\documentstyle[12pt]{article}
\begin{document}

\begin{center}
\begin{Large}
{\bf Thermodynamics of Heat Shock Response} 
\end{Large}
\vspace{1.0cm}

\begin{large}
Kristine Bourke Arnvig$^a$, Steen Pedersen$^a$ and Kim Sneppen$^b$
\end{large}
\vspace{0.5cm}

$^a$ Institute of Molecular Biology, {\O}ster Farimagsgade 2A,
1353 Copenhagen K, Denmark.\\
$^b$ Nordita, Blegdamsvej 17, 2100 Copenhagen {\O}, Denmark
\vspace{0.5cm}

July 26, 1999
\end{center}
\vspace{1.0cm}

\begin{small}
\noindent
{\sl Abstract:} Production of heat shock proteins
are induced when a living cell is exposed to 
a rise in temperature.
The heat shock response of protein DnaK
synthesis in {\sl E.coli} for temperature shifts 
$T\rightarrow T + \Delta T$  and
$T\rightarrow T - \Delta T$  is measured
as function of the initial temperature $T$. 
We observe a reversed heat shock at low $T$.
The magnitude of the shock increases when
one increase the distance to the temperature $T_0 \approx 23^o$,
thereby mimicking the
non monotous stability of proteins at low temperature.
This suggest that stability related to hot as well as
cold unfolding of proteins is directly implemented in the
biological control of protein folding.
\end{small}
\vspace{0.5cm}

\vspace{1.0cm}

Chaperones direct protein folding in the living cell by
binding to unfolded or misfolded proteins. The expression 
level of many of these catalysts of protein folding change 
in response to environmental changes.  In particular, 
when a living cell is exposed to 
a temperature shock the production of these proteins
are transiently increased. The response is seen for all organisms
\cite{Lindquist}, and in fact involves related proteins across
all biological kingdoms. 
The heat shock response (HS) in {\sl E.coli} 
involves about 40 genes that 
are widely dispersed over its genome \cite{Neidhardt,Gross}.
For {\sl E.coli} the response is activated through
the $\sigma^{32}$ protein. 
The $\sigma^{32}$ binds to RNA polymerase (RNAp)
where it displaces the $\sigma^{70}$ subunit 
and thereby changes RNAp's affinity to a number of promoters
in the {\sl E.coli} genome. This induce production 
of the heat shock proteins.
Thus if the gene for $\sigma^{32}$ 
is removed from the {\sl E.coli} genome the HS is suppressed 
\cite{Neidhardt,Zhou}
and also the cell cannot grow above $20^o C$.

The HS is fast. In some cases it can be detected by a changed 
synthesis rate of {\it e.g.} the chaperone protein DnaK 
already about a minute after the temperature shift.
Given that the DnaK protein in itself takes about 45 seconds 
to synthesize the observed fast change in DnaK production 
must be very close to the physical mechanism that trigger the response.
In fact we will argue for a mechanism which does
not demand additional synthesis of $\sigma^{32}$ 
in spite of the fact that DnaK is only expressed
from a $\sigma^{32}$ promoter, and thus postulate
that changed synthesis of $\sigma^{32}$ only plays a role
in the later stages of the HS.
To quantify the physical mechanism we measure the dependence
of HS with initial temperature and find
that the magnitude of the shock is inversely proportional to the
folding stability of a typical globular protein.
\vspace{0.5cm}

The present work measure the expression of protein DnaK.
Steady state levels at various temperature and growth
conditions can be found in \cite{Herendeen,Pedersen}.
The steady state number of DnaK in an {\it E.coli} cell varies from 
approx. 4000 at $T=13.5$ to approx. 6000 at $37^oC$,
thereby remaining roughly proportional to number of ribosomes.
DnaK is a chaperone, and have a high affinity for hydrophobic
residues \cite{Gragarov}, as these signal a possible misfold
(for folded proteins the hydrophobic residues are typically
in the interior).
$\sigma^{32}$ control the expression of DnaK, 
the $\sigma^{32}$ must bind to the RNAp before
this bind to the promoter for DnaK.
One expect at most a few hundred $\sigma^{32}$ in the cell, a number which
is dynamical adjustable because
the in vivo half life of $\sigma^{32}$ is short
(in steady state it is 0.7 minutes at $42^o$C and $15$ minutes at $22^o$C
\cite{Straus,Tilly89}). The life time $\sigma^{32}$ is known
to increase transiently under the HS.

The measurement was on {\sl E.coli} K12 strain
grown on A+B medium \cite{Reeh} with a $^3H$ labelled amino acid added.
After the temperature shift we extracted 0.6 ml samples of the culture 
at subsequent times. Each sample was exposed to radioactive labelled 
methionine for 30 seconds, 
after which non radioactive methionine was added in huge excess
($10^5$ fold).  Methionine is an amino acid that the bacteria
absorb very rapidly, and then use in protein synthesis.
Protein DnaK was separated by
2-dim gel electrophoresis as described by O'Farrell \cite{Farrell},
and the amount of synthesis during the 30 seconds
of labeled methionine exposure
was counted by radioactive activity
and normalized first with respect
to $^3H$ count and then with respect to total protein production.
This result in an overall accuracy of about $10\%$.
The result is a count of the differential rate
of DnaK production (i.e. the fraction DnaK constitute of
total protein synthesis relative to the same fraction before
the temperature shift \cite{Reeh})
as function of time after the temperature shift.
For the shift $T\rightarrow T+\Delta T$ at time $t=0$
we thus record:
\begin{equation}
r(T,t) \;=\; \frac{Rate \;\; of \;\; DnaK\;\; 
production\;\; at\;\; time\;\; t}
{Rate \;\; of \;\; DnaK\;\; production \;\; at \;\; time\;\; t=0 }
\end{equation}
where the denominator counts steady state
production of DnaK at temperature $T$.
In figure 1 we display 3 examples,
all associated to temperature changes of 
absolute magnitude $\Delta T=7^oC$.
When changing $T$ from $30^oC$ to $37^oC$
one observe that $r$ 
increases to $\sim 6$ after a time of 0.07 generation. 
Later the expression rate relaxes to normal level again, 
reflecting that other processes counteracts the initial response.
When reversing the jump, we see the opposite effect, 
namely a transient decrease in expression rate.
Finally we also show a temperature jump at a low temperature,
and here we observe the opposite effect, namely that a $T$ increase
give a decrease in expression rate. Here a corresponding
$T$ decrease in fact
gives an increase in expression rate (not shown).
Thus the cells response to a positive temperature jump is
opposite at low temperature $T$  than it is at high $T$.

In figure 2 we summarize our findings by plotting 
for a number of positive shocks $T\rightarrow T+7^oC$
the value of $r=R$ where the deviation from $r=1$ is largest. 
This value can be well fitted by the following dependence
on temperature $T$
\begin{equation}
ln ( R (T) )\;=\; ( \alpha \Delta T) \; ( T - T_0 ) 
\end{equation}
where $R(T = T_0 = 23^o) = 1$ 
and $\alpha \Delta T \;=\; \frac{ln(R_1/R_2)}{T_1-T_2}$
= $0.2\cdot K^{-1}$ (i.e. $\alpha=0.03 K^{-2}$).

In order to interpret this result we first assume that
the production rate of DnaK is controlled by two factors,
a slowly varying factor $C$ 
that depends on composition
of some other molecules in the cell, and an instantaneous
chemical reaction constant $K$. 
Thus at time $t$ after a shift in temperature 
the production of DnaK in the cell is:
\begin{equation}
\frac{d [DnaK ]}{dt} (t,T\rightarrow T+\Delta T) 
\;=\; C(t,T\rightarrow T+\Delta T)
\; \cdot K(T+\Delta T)
\end{equation}
where the initial composition of molecules,
$C(t=0,T\rightarrow T+\Delta T)$ equals
their equilibrium number at the temperature we changed from, i.e.
$=C_{eq}(T)$. 

To lowest approximation, where we even
ignore feedback from changed DnaK in the cell
until DnaK production rate have reached its peak value:
\begin{equation}
R \;=\; \frac{K(T+\Delta T)}{K(T)}
\end{equation}
which implies that
\begin{equation}
ln(R) \;=\; ln(K(T+\Delta T))-ln(K(T)) \;=\; \frac{dln(K)}{dT} \Delta T
\end{equation}
Using the linear approximation in Fig. 2 
\begin{equation}
ln(K)\; \approx \; const \; +\; \frac{\alpha}{2} (T-T_0)^2
\end{equation}
Identifying $K=exp(-\Delta G/T)$ the 
effective free energy associated to the reaction is
\begin{equation}
\Delta G \;\approx \; G_0 \; -\; \frac{\alpha T}{2} (T-T_0)^2
\end{equation}
Thus $\Delta G$ has a maximum at $T=T_0=23^o$.
\vspace{0.5cm}

To interpret the fact that HS is connected
to a $\Delta G$ that have a maximum at $T=T_0\approx 23^o$
we note that many proteins exhibit a maximum stability 
at $T$ between $10^o$C and $30^o$C \cite{Priv89,Priv95}.
Thus $\Delta G=G(folded)-G(unfolded)$ connected to the 
folded state of a protein is at a minimum at $T_0$.
The corresponding maximum of stability is in effect 
the result of a complicated balance between
destabilization from entropy of polymer degrees of freedom
at high $T$, and destabilization due to 
decreased entropic contribution to hydrophobic
stabilization of proteins at low $T$ \cite{Priv95,Hansen}.
One should expect a similar behaviour also for some parts 
of a protein \cite{Hansen}, and thus expect a max binding
for hydrophobic protein-protein associations around $T_0$.
Quantitatively the size of the $\Delta G$ change 
inferred form the measured value of $\alpha=0.03K^{-2}$
correspond to a changed $G$ of about $20\rightarrow 30$kT 
(about 15Kcal/mol),
for a temperature shift of about 40-50$^oC$.
This matches the change observed for typical 
single domain proteins \cite{Priv95}. 
Thus the HS is associated to a $\Delta G$ change
equivalent to the destabilization of a typical protein.
\vspace{0.5cm}

The above picture still leave us with the puzzle that
protein binding and folding stability
is at a maximum around $T_0$,
whereas the effective $\Delta G$ we observe have a minimum there.
This can only be reconciled if the interaction we consider
is inhibitory.  An inhibitory binding that
controls the feedback is indeed possible \cite{Gross}.
To summarize our understanding we in
figure 3 display the molecular network that we believe is controlling
the transient heat shock levels of DnaK in the cell.
The key inhibitory control mechanism is the association
of DnaK to $\sigma^{32}$. 
DnaK binds to unfolded protein residues \cite{Gragarov},
and the amount of DnaK$ \cdot \sigma^{32}$ association
thereby monitor cellular consequences of a shift in temperature.

Impacts of mutants: We have measured the heat shock in
a strain where the $\sigma^{32}$ gene 
is located on a high copy
number plasmid. In this strain where the synthesis rate for 
$\sigma^{32}$ may
approach that of DnaK we find a HS that was smaller 
and also remained positive down to temperature
jumps from $T$ well below $T_0=23$.
According to fig. 3 this reflects a situation where both $\sigma^{32}$
and DnaK are increased. 
With increased DnaK level, one may have a situation where
DnaK exceeds the amount of unfolded proteins, and 
free DnaK concentration thus becomes nearly 
independent of the overall state of proteins in the cell.
Also the huge increase in $\sigma^{32}$ supply 
may decrease the possibility for the sink to act effectively.
Thereby other effects as {\it e.g.} 
the temperature dependence of the 
binding $\sigma^{32} \cdot RNAp$ 
versus the binding $\sigma^{70} \cdot RNAp$ 
(i.e. $K_{32}/K_{70}$ from figure 3)
may govern a response
that otherwise would be masked by a 
strongly varying inhibition from $\sigma^{32}$ binding to DnaK.
\vspace{0.5cm}

The reaction network in Figure 3 allow a more careful 
analysis of the production rate of DnaK:
\begin{equation}
\frac{d [ DnaK ] }{dt} \;\; \propto \;\; [RNAp \cdot \sigma^{32}]
\approx
\frac{ [ \sigma^{32} ]}{1 \; +\;  g \; [ DnaK ]}
\end{equation}
where the $\sigma^{32}$ changes
when bound to DnaK due to degradation by proteases 
(the ``sink'' in figure 3):
\begin{equation}
\frac{d[ \sigma_{32} ]}{dt} \;\; \propto \;\; 
Supply \; -\; [DnaK \cdot \sigma^{32}] \;\;
\approx \;\;
Supply \;\; - \;\; \frac{[\sigma^{32}]}{1+(g \;  [ DnaK ])^{-1}}
\end{equation}
Here $g=exp(-\Delta G/T)$ is an effective reaction constant.
In the approximation where we ignore free $\sigma^{32}$, free $\sigma^{70}$
and the fraction of DnaK bound by $\sigma^{32}$ then:
\begin{equation}
g \; =\; \left( \frac{K_{70} [\sigma^{70}]}{K_{32} [RNAp]} \right)
\cdot \left( \frac{K_{D32}}{1+K_{DU} [U_f]} \right) 
\end{equation}
The first term expresses the $\sigma$'s competition for RNAp binding
whereas the second term expresses the DnaK controlled response. 
$[U_f]$, which denote unfolded proteins that are not bound to DnaK, 
decreases with increasing [DnaK].

When moving away from $T_0$, 
i.e. lowering $g$ by increasing $[U_f]$,
the rate for DnaK production increases.
For an approximately unchanged ``Supply'' the extremum in production
occurs when $d [\sigma^{32}]/dt=0$ and has a value 
that approximately is $\propto \; 1/g$.
With the assumption that ``Supply" does not have time
to change before extreme response is obtained,
we identify $R$ with $1/g$ and thereby with 
the free energy difference $\Delta G$ that controls the HS.
The early rise in $r$ is reproduced when most $\sigma^{32}$ are bound to 
RNAp reflected in the condition $g [DnaK]<<1$.
This implies a significant
increase in $\sigma^{32}$ lifetime under a positive HS,
and implies that the early HS is due to a changed
depletion rate of $\sigma^{32}$. Later the response is modified,
partly by a changed ``Supply'' and finally by a changed
level of the heat shock induced protease HflB that depends 
on and counteracts the $\sigma^{32}$ level in the cell.

The largest uncertainty in our analysis is the possibility
of a significant time variation in ``Supply'' and HflB 
level during the HS.
As these will govern the late stages of the heat shock, in
particular including its decline, the variation in 
``$\Delta G$'' for proteins in the cell may easily be 
underestimated from using the peak height variation with $T$.
Adding to the uncertainty in what $\Delta G$ precisely represents
is also the fact that although we only measure DnaK, 
it can be the complex of the heat shock proteins
DnaK, GrpE and DnaJ that sense the state of unfolded proteins in the cell,
as removal of any of these display an increased life time of $\sigma^{32}$
\cite{Straus90,Tilly89}. Such cooperativity may amplify
the heat shock. 

For the final interpretation of $\Delta G$ 
we stress that it effectively counts the free 
energy difference between 
the complex DnaK$\cdot \sigma^{32}$ and that of 
DnaK being free or being bound to unfolded proteins in the cell.
Dependent on the fraction of DnaK relative to unfolded proteins [U] 
in the cell, i.e. whether $K_{DU}[U_f]$ is larger or smaller than 1,
the HS will depend or not depend on the 
overall folding stability of proteins in the cell.
Thus for much more unfolded proteins than DnaK in the cell,
the measured $\Delta G$ reflect both an 
increases of the binding to unfolded residues 
$K_{DU} [U_f]$ as well as a decrease of the
DnaK$\cdot \sigma^{32}$ binding $K_{D32}$ when moving away from $T_0$.
Our data does not discriminate between these processes.
This discrimination can however be obtained from the data of ref.
\cite{Tilly83} where it was found that
overexpression of DnaK through a $\sigma^{32}$ dependent pathway in fact
repress HS. 
As DnaK$\cdot \sigma^{32}$ binding still play a crucial 
role in this setup, the vanishing HS of \cite{Tilly83} 
support a scenario where too much DnaK imply $K_{DU} [U_f]<<1$. 
Then $g$ in eq. 10 and thereby the $\sigma^{32}$
response becomes insensitive to
the amount of unfolded proteins in the cell.

We conclude that the HS is induced through the changed 
folding stability of proteins throughout the cell, 
sensed by a changed need of chaperones. 
We believe this reflect primarily an increased amount of 
proteins that are on the way to become folded (nascent proteins),
and not an increased denaturation of already folded proteins,
because spontaneous denaturation of proteins 
is extremely unlikely at these temperatures.
Thus the deduced sensitivity to the thermodynamic stability
of proteins may primarily reflect a 
correspondingly sensitivity to change in folding times.
\vspace{0.5cm}

We now discuss related proposals of
``cellular thermometers'' for the HS.
McCarthy et al. \cite{McCarthy88}
proposed that the thermometer was a change in
autophosphorylation of the DnaK protein. 
This should cause a temperature dependent activity of this protein. 
However their data does not indicate
that the reversed HS response that we observe at $T<23$°C 
could be caused by such mechanism. 
Gross \cite{Gross} made an extensive network of possible chemical
feedback mechanisms which connect
a rise in $\sigma^{32}$ level with
the folding state of proteins in the cell.
It included HS induced through 
an increased synthesis of $\sigma^{32}$,
an increased release of $\sigma^{32}$ from DnaK
as well as an increased stability of $\sigma^{32}$ 
when DnaK gets bound to unfolded protein residues.
Our fig. 3 specifies these possibilities
to a minimalistic chemical response including the two latter
mechanisms combined, and of these only the option
of a changed stability of $\sigma^{32}$ 
due to a sink controlled by DnaK$\cdot \sigma^{32}$ 
is able to reproduce also the fact that the max HS
takes time to develop.
In regards to the by Morita \cite{Morita} proposed 
increased synthesis of $\sigma^{32}$,
we note that for high temperatures $T$, 
the major mechanism that controls $\sigma^{32}$ synthesis
in fact is a $T$ dependent change in the mRNA structure that
leads to an increased translation at
increased $T$ \cite{Morita}. 
However again our finding of a reversed HS at $T<23$°C is not
readily explained by such changes in the stability of mRNA structures
below 23°C.  
\vspace{0.5cm}

In summa, we observed that positive heat shock is induced 
when $T$ changes away from $T_0\sim 23^o$. 
We found that the size of the heat shock qualitatively as well
as quantitatively follows the thermodynamic stability of proteins
with temperature. 
This suggested that stability related to hot as well as 
to cold unfolding of proteins is implemented the in HS.
We demonstrated that such an implementation was possible
in a minimalistic chemical network 
where the control is through an inhibitory binding of the
central heat shock proteins.
We finally saw that the temporal behaviour
of the HS is reproduced when 
this inhibitory binding controls the heat shock 
by exposing $\sigma^{32}$ to a protease.

{\bf Figure Captions}

\begin{itemize}

\item Fig. 1. Heat shock response measures as DnaK rate production change
as function of time since a temperature shift. 
The production is normalized with overall protein production rate,
as well as with its initial rate.
In all cases we use absolute $\Delta T=7^oC$.
We stress that the time scale is in units of one bacterial generation
measured at the initial temperature.
At $T=37^o$C the generation time is 50 min,
at $30^oC$ it is 75 min and at $20^o$C it is 4 hours.

\item Fig. 2. Induction fold $R$ for positive temperature jumps
as function of initial temperature. The straight line correspond to the
fit used in eq. 2.

\item Fig. 3. Sufficient molecular network for the 
early heat shock.
All dashed lines with arrows in both ends are chemical reactions
which may reach equilibrium within a few seconds 
(they represent the homeostatic response).
The full directed arrows represent one way reactions, with the
production of DnaK through the $\sigma^{32} \cdot RNAp$ complex being
the central one in this work (this step is catalytic, 
it involves DNA translation etc.) 
The time and temperature dependence of the early HS
is reproduced when most DnaK is bound to unfolded proteins,
and when remaining DnaK
binds to $\sigma^{32}$
to facilitate a fast depletion of $\sigma^{32}$ 
through degradation by protease HflB. 
\end{itemize}


\begin{thebibliography}{10}

\bibitem{Lindquist}
S. Lindquist, {\it Annu. Rev. Biochem.} {\bf 55}, 1151 (1986).

\bibitem{Neidhardt}
F.\ C. Neidhardt, R.\ A. VanBogelen and V. Vaughn,
{\it Annu. Rev. Genet.} {\bf 18}, 295 (1984).

\bibitem{Gross}
C.A. Gross, Cellular and Molecular Biology, ASM press, 1382 (1996).

\bibitem{Zhou}
Y. Zhou, N. Kusukawa, J.\ W. Erickson, C.\ A. Gross
\&  T. Yura, {\it Journ. of Bacteriology} {\bf 170}, 3640 (1988).

\bibitem{Herendeen}
S.\ L. Herendeen, R.\ A. VanBogelen and F.\ C. Neidhardt,
{\it J. Bacteriol.} {\bf 139}, 185 (1979).

\bibitem{Pedersen}
S. Pedersen, P.\ L. Bloch, S. Reeh and F.\ C. Neidhardt,
{\it Cell} {\bf 14}, 179 (1978).

\bibitem{Gragarov}
A. Gragarov, L. Zeng, X. Zhao, W. Burkholder and M.\ E. Gottesman
{\it J. Mol. Biol.} {\bf 235}, 848 (1994).

\bibitem{Straus}
D.\ B. Straus, W.\ A. Walter and C.\ A. Gross,
{\it Nature} {\bf 329}, 348 (1987).

\bibitem{Straus90}
D.\ B. Straus, W.\ A. Walter and C.\ A. Gross,
{\it Genes Dev.} {\bf 4}, 2202 (1990).

\bibitem{Tilly89}
K. Tilly, J. Spence and C. Georgopoulos,
{\it Journ. of Bacteriology} {\bf 171}, 1585 (1989).

\bibitem{Farrell}
P.H. O'Farrell, {\it J. Biol. Chem} {\bf 250}, 4007 (1975).

\bibitem{Reeh}
S. Reeh, S. Pedersen and J. D. Friesen,
{\it Molec. gen. Genet.} {\bf 149}, 279 (1976).

\bibitem{Priv89}
P.\ L. Privalov, E.\ I. Tiktopulo, S.\ Yu. Venyaminov,
Yu.\ V., Griko, G.\ I. Makhatadze  \&
N.\ N. Khechinashvili, {\it J. Mol. Biol.} {\bf 205}, 737 (1989).

\bibitem{Priv95}
G.\ M. Makhatadze and P.\ L. Privalov,
{\it Adv. Protein Chem.} {\bf 47}, 307 (1995)

\bibitem{Hansen}
A. Hansen, M.\ H. Jensen, K. Sneppen and G. Zocchi, 
{\it Eur. Phys. J B} {\bf 6}, 157 (1998).

\bibitem{Tilly83}
K. Tilly, N. McKittrick, M. Zylicz and C. Georgopoulos,
{\it Cell} {\bf 34}, 641 (1983).

\bibitem{McCarthy88}
J.\ S. McCarthy and G.\ C. Walker, {\it Proc. Natl. Acad. Sci} {\bf 88},
9513 (1991)

\bibitem{Morita} 
M.\ T. Morita, Y. Tanaka, T.\ S. Kodama, Y. Kyogoku,
H. Yanagi and T. Yura, {\it Genes \& Dev.} 
{\bf 13}, 655 (1999).

\end{thebibliography}
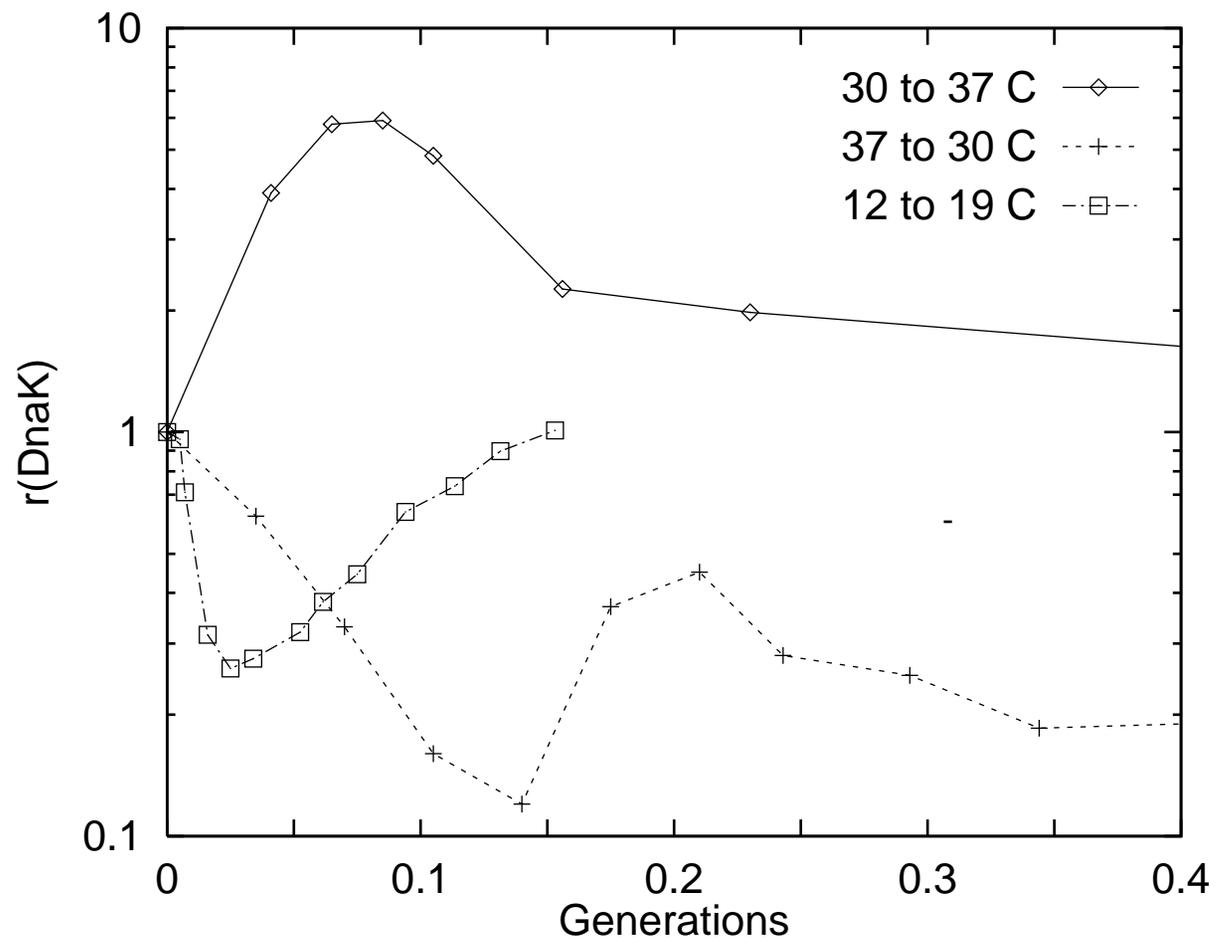
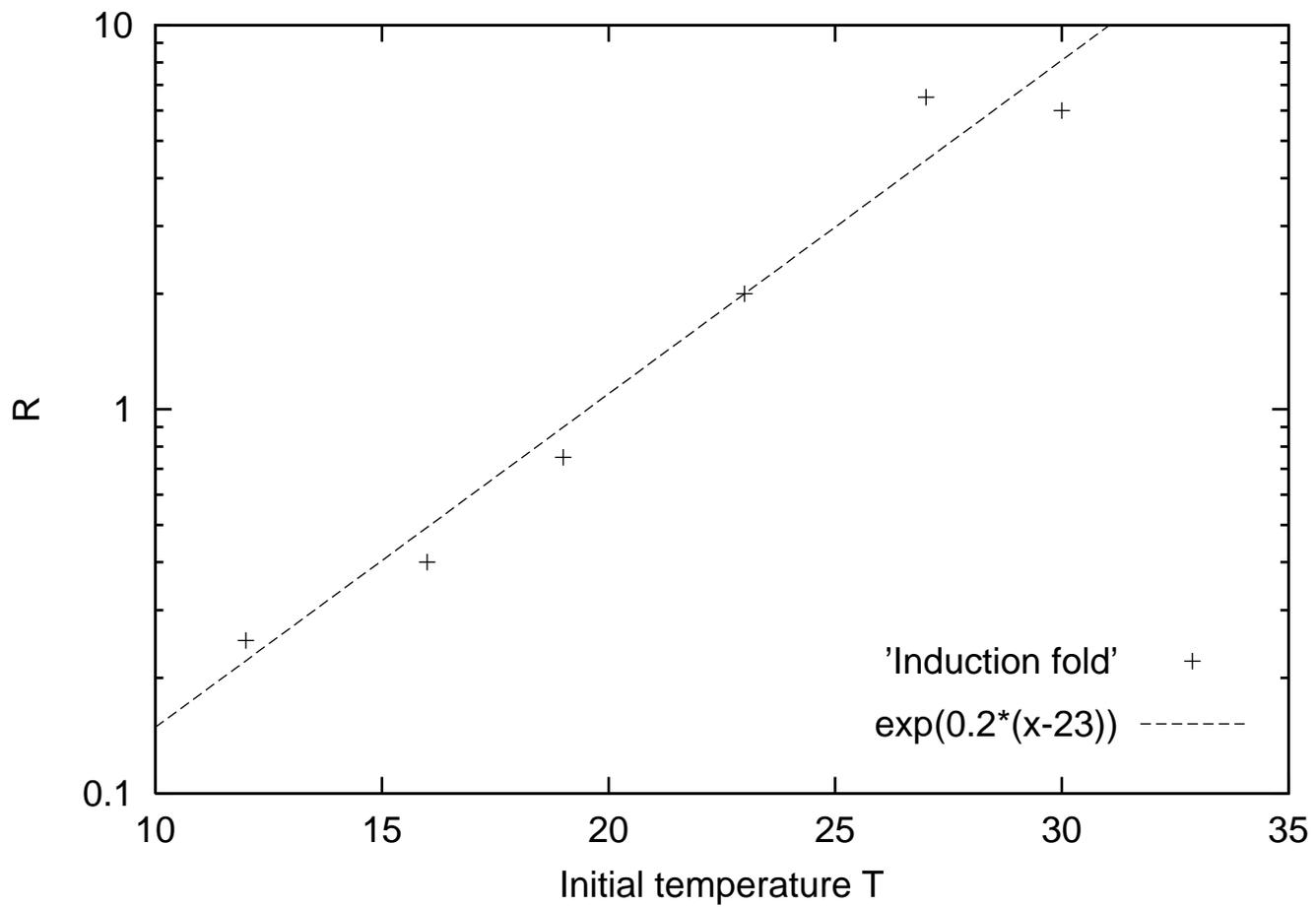
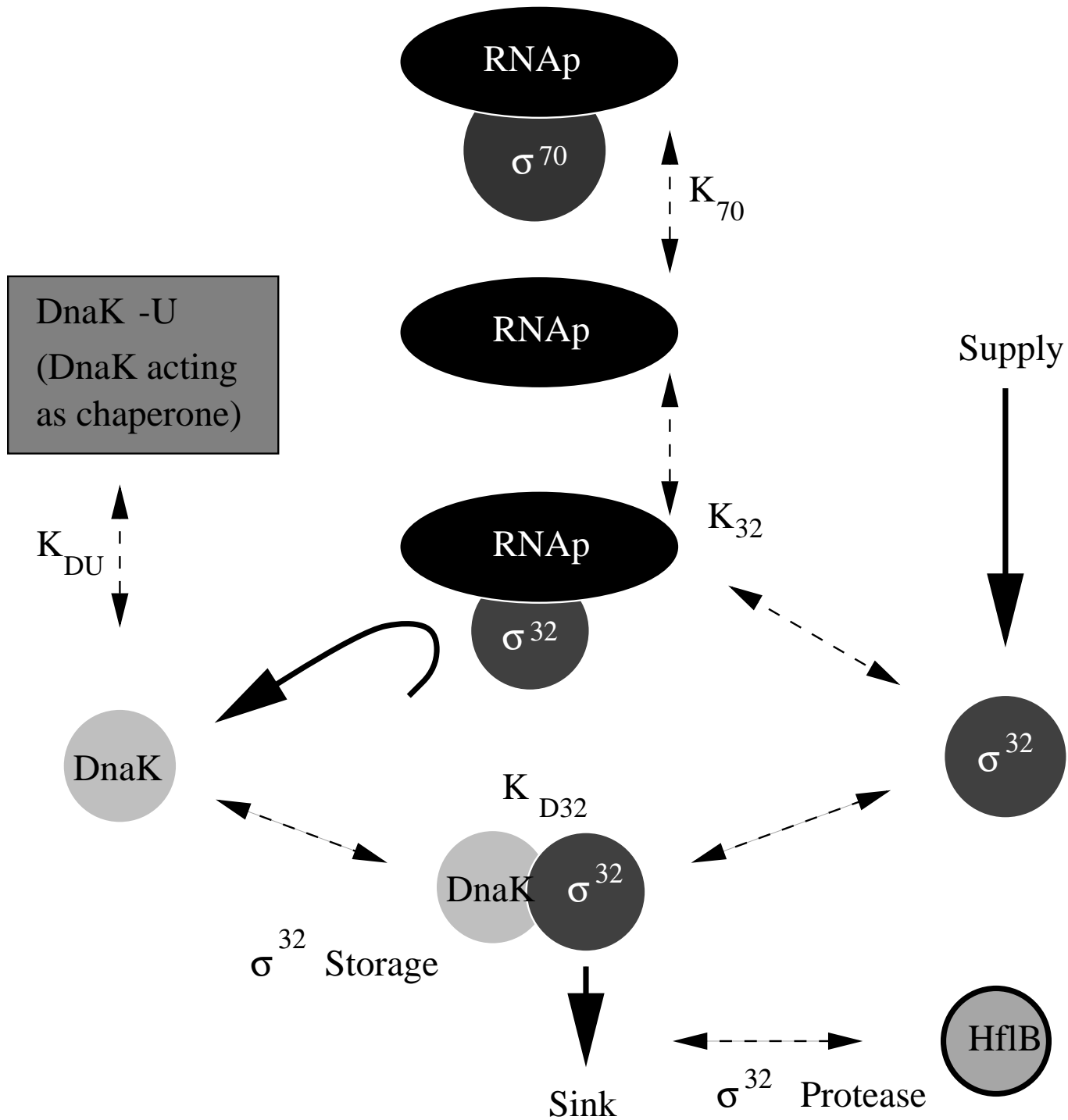
\end{document}